\documentclass[12pt]{iopart}
\usepackage{iopams}  
\usepackage{epsfig}
\begin{document}

\title[]{Results from the LISA Phase Measurement System
  Project\footnote[7]{ESA/ESTEC contract 15658/01/NL/EC, Technical
    Officer Alberto Resti}}

\author{David Summers\dag\footnote[3]
{To whom correspondence should be addressed (David.Summers@sea.co.uk)}
and David Hoyland\ddag  
}

\address{\dag\ Systems Engineering \& Assessment Ltd, SEA House, PO Box 800,
Fishponds, Bristol BS16 1SU, UK}

\address{\ddag\ University of Birmingham School of Physics and Astronomy,
Edgbaston, Birmingham B15 2TT, UK}

\begin{abstract}
This article presents some of the
more topical results of a study into the
LISA phase measurement system. This system is responsible for
measuring the phase of the heterodyne signal caused by the
interference of the laser beams between the local and far spacecraft.
Interactions with the LISA systems that surround the phase measurement system
imply additional non-trivial requirements on the phase measurement system.
\end{abstract}



%\maketitle

\section{Introduction}
Systems Engineering \& Assessment Ltd., in association with the
Universities of Birmingham and Glasgow, has carried out a
study into the Phase Measurement System that will be required
to measure the optical heterodyne signals on board the LISA
spacecraft. The LISA mission forms one of the most ambitious space missions
ever conceived, where many difficulties have to be overcome using
novel techniques, and tightly coupled systems. The interactions
between the various LISA systems puts very strong requirements on the
various systems, and the Phase Measurement System in particular.

This reports summarises some of the more topical drivers for the
design of the Phase Measurement and associated systems.

\section{Time Delay Interferometry}

\begin{figure}
\begin{center}
\psfig{figure=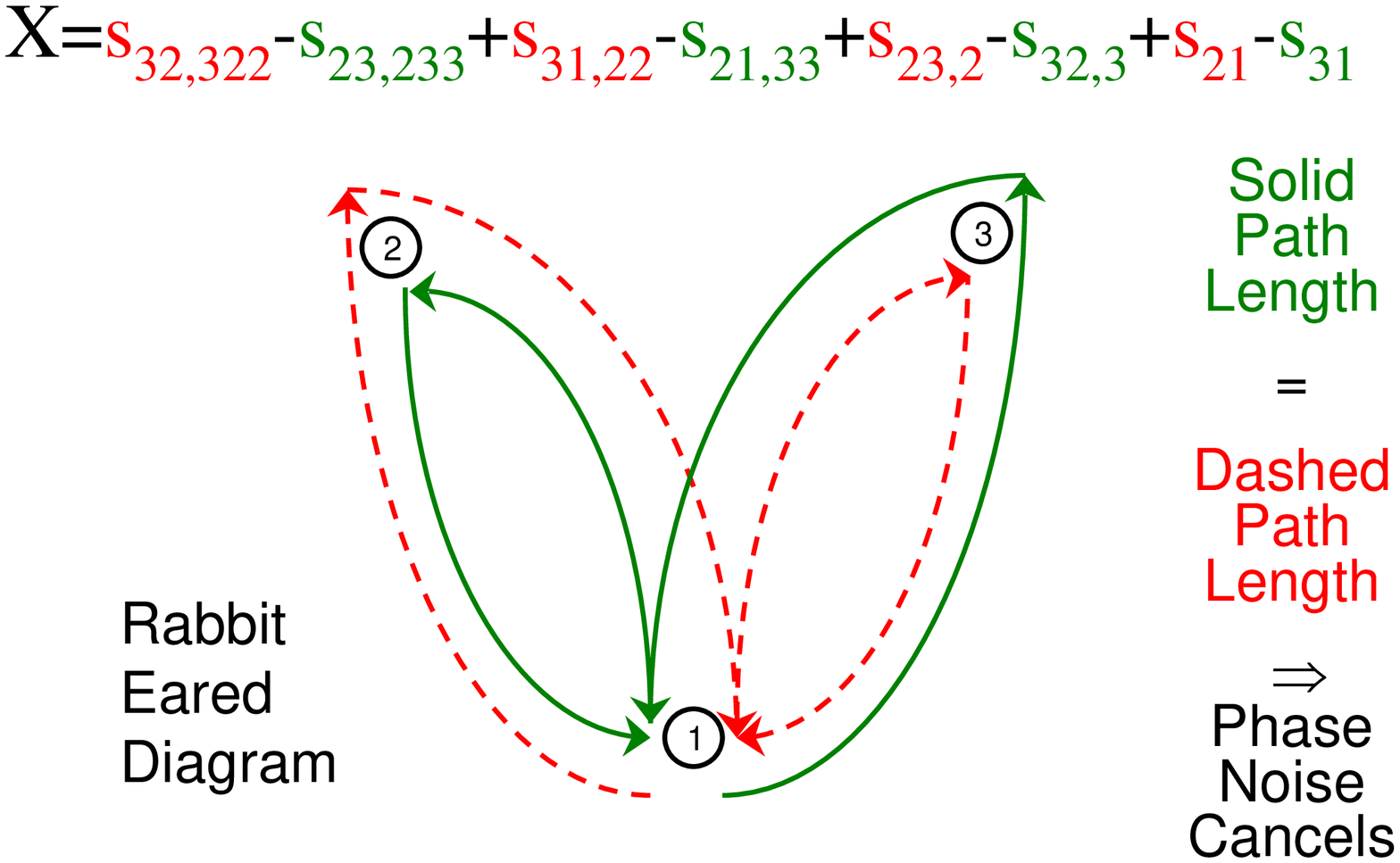,width=8cm}
\end{center}
\caption{\label{fig:rabbit}The phase measurements that make up the TDI X variable}
\end{figure}

One of the strongest drivers on the Phase Measurement System design is
the requirement to cancel laser phase noise by approximately seven orders
of magnitude. The baseline method for
cancelling laser phase noise is Time Delay Interferometry (TDI)
\cite{TDI}. This technique cancels the phase noise through
combinations of time delayed measurements. To give an example of how
these cancellations occur consider a simplified version of the TDI1 X
variable. This is illustrated graphically in the Rabbit Eared Diagram
\Fref{fig:rabbit} \cite{algo}. 

The phase measurements give the difference in phase between the local
laser {\it at the current time} and the distant laser {\it at a prior
time}. By combining with the phase measurements at another spacecraft, taken at
earlier times the distant laser phase noise is cancelled, at
the expense of introducing phase noise at earlier times at yet another
laser. By combining two phase streams, shown as the solid and dashed lines
in \Fref{fig:rabbit}, after four measurements each the phase noise for the
two data steams on spacecraft 1 are identical. This means that
subtracting the two streams the X variable cancels the laser
noise. In the X variable the two data streams are distinguished by the
sign of each term, either summing or subtracting.

Effectively the TDI variables construct an equal arm
interferometer. Instead of the laser beams being reflected TDI
utilises phase measurements to construct the two equal length
arms. This means that the intuition developed for 
equal arm interferometers is equally applicable to LISA with TDI.

In particular the difference in arm lengths introduces the laser
frequency noise into the TDI variable, in terms of the Laplace transform:
\begin{equation}
N_{\phi}=N_f (1-\exp (- \Delta t s ) )/s \simeq N_f
\Delta t \qquad \mbox{(for $s \ll 1/\Delta t$)}
\end{equation}
where $N_f$ is the laser frequency noise, and $\Delta t$ the
time difference between the arm lengths. For laser frequency noise at
the level of $30{\rm Hz/\sqrt{Hz}}$ this implies that the arm lengths
need to be equal within 10m or 33ns in order for the phase noise to be
kept below $10^{-6} {\rm cycles/\sqrt{Hz}}$. This requirement also
applies to the synchronisation of the various phase measurements inside
the TDI variables.

This has several implications on the phase measurement system:
\begin{itemize}
\item The phase measurements time stamp need a fidelity of
  ${\cal O}(10 \rm ns)$.
\item The phase measurements on independent spacecraft needs to be
  synchronised to ${\cal O}(10 \rm ns)$.
\item Multiple Phase Measurement Systems will be required per
  heterodyne, to ensure than several phase measurements can be made on
  a single heterodyne, but with {\it precise} and {\it varying} time
  separation.
\end{itemize}
The synchronisation of phase measurements on distinct spacecraft is
especially onerous, as it apparently requires an entirely new LISA
system that synchronises the spacecraft, either:
\begin{itemize}
\item Absolutely synchronising the spacecraft clocks to 
  ${\cal O}(10 \rm ns)$ {\it and} knowledge of the spacecraft
  separation to ${\cal O}(10 \rm m)$.
\item Or synchronising the phase measurements along null light like
  vectors between the spacecraft.
\end{itemize}
Although recent work \cite{interpol} suggests that interpolation
between phase measurements may also provide the required
synchronisation without the need for a Synchronisation System.

\section{Decimation}

A typical phase measurement system forms an average of the phase
difference over the phase measurement time period.
\begin{equation}
\label{decim}
M(\Delta\phi,t)=\frac{1}{\Delta t} \int_{t-\Delta t}^t dt'
\Delta\phi(t') 
= \frac{1}{\Delta t} \int_{t-\Delta t}^t dt'\phi_1(t')
-  \frac{1}{\Delta t} \int_{t-\Delta t}^t dt'\phi_2(t')
\end{equation}
Where measurements take place at the time intervals
$M(\Delta\phi,n\Delta t)$ for $n=0,1,\ldots$.  Considering equation
\ref{decim} as a continuous time process the suppression of phase noise
falls as $1/s$ in the high frequency limit. For laser frequency noise
at the level of $N_f$ this gives the phase measurement process $M$
noise which falls as:
\begin{equation}
N_M=\frac{2\pi N_f}{\Delta t s^2}
\end{equation}
This only falls slowly with increasing frequency ($\sim 2\pi s$), and this
implies significant aliasing of the noise into the decimated phase
measurement. In order for the noise to be below 
$10^{-6}\hbox{ cycles}/\sqrt{\rm Hz}$ at frequencies above the decimation
Nyquist Frequency ($1/(2\Delta t)$), and for laser frequency noise at
the level $30{\rm Hz}/\sqrt{\rm Hz}$, the sampling rate has to be in
excess of $\Delta t=1 \mu\rm s$. Such a high rate is infeasible for a
space mission such as LISA because of the unacceptably high telemetry
rate it would imply. This has the possible implications:
\begin{itemize}
\item The aliasing of phase noise into the phase measurement in
  equation \ref{decim} although severe, is linear in the laser phase
  noise. This means that at distinct phase measurements, {\it e.g.} on
  separate spacecraft, that the aliasing of phase noise is identical
  for all laser phase noise. Hence the aliasing of phase noise just
  increases the level of the noise, and does not invalidate the TDI
  methodology. So it is possible to accept some aliasing of the
  phase noise, but at the expense of increasing the level of
  synchronisation needed to cancel the noise.
\item The phase measurement process $M$ needs to be designed to
  significantly suppress the high frequency noise to avoid aliasing
  problems. This can be achieved using analogue filters, digital
  filters on the phasor signal (such as Fourier Transform windowing
  functions), and digital filters on the extracted phase. This
  significantly increases the design complexity of the digital system.
  In order to avoid the aliasing of noise the signal needs to be
  over sampled, this implies overlapping digital processing, as for
  example is familiar when windowing is used in FFT applications. 
\end{itemize}
It is expected that a combination of these will be ultilised in LISA,
and its is necessary to ensure that the LISA methodology is compatible.

\section{Phase Noise Measurements}

The level of laser phase noise has another major impact on the phase
measurement system, in that the heterodyne signal being measured is
intrinsically noisy. The Phase Measurement System must express this
noise in a form where it can be cancelled by phase differences used in
TDI \cite{LTP,task3,task4}.
 
\begin{figure}
\begin{center}
\psfig{figure=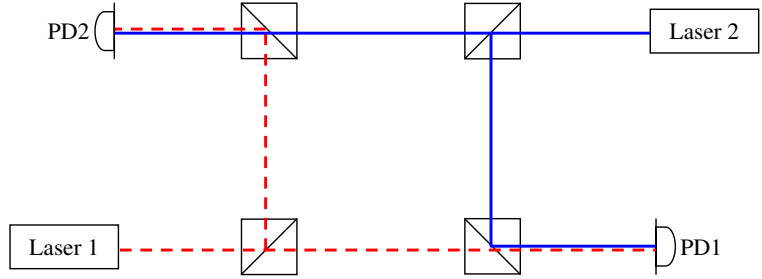,width=10cm}
\end{center}
\caption{\label{fig:gedanken}Gedankenexperiment that illustrates the
  phase noise cancellations}
\end{figure}

The simplest configuration that illustrates the difficulty is shown in
\Fref{fig:gedanken}. In this gedankenexperiment all components are
mounted on a stable optical bench; the lasers each have
independent phase noise. The two photodiodes each measure a noisy
signal, caused by the laser phase noise. However the phase difference
between the photodiodes is a property of the optical bench and hence
should be constant.

This can be understood as arising because two independent frequencies
of phase noise contribute to noise at the measurement frequency, both
baseband and at twice the frequency. The noise at baseband
cancels in differences between phase measurements, as is
required by TDI; however the noise at twice the frequency is more
complex. Consider a signal with a small amount of phase noise at
twice the main frequency:
\begin{equation}
\cos(\omega t + \phi + \epsilon \sin( 2 \omega t))
\simeq \cos(\omega t + \phi) - \frac12\epsilon \cos(\omega t - \phi)
+ \frac12\epsilon \cos(3 \omega t + \phi)
\end{equation}
The phase noise is responsible for a second term $\cos(\omega t -
\phi)$ at {\it exactly} the heterodyne frequency, {\it i.e.} it can not be
removed from the signal. If this signal is measured by a phase meter
which converts the heterodyne to a phasor, {\it e.g.} a Fourier
Transform based phase measurement techniques, then:
\begin{equation}
\cos(\omega t + \phi + \epsilon \sin( 2 \omega t))
\stackrel{FT}\Rightarrow \rme^{i \phi} - \frac12 \epsilon \rme^{-i \phi}
\end{equation}
Unfortunately the $\rme^{-i \phi}$ term does not cancel in phase
differences. Consider the phase measurement between two heterodynes
with phases $\phi_1$ and $\phi_2$, with common phase noise. The phasor
for the measured phase difference is:
\begin{equation}
\hskip -1cm
\frac{\rme^{i \phi_1} - \frac12 \epsilon \rme^{-i \phi_1}}
{\rme^{i \phi_2} - \frac12 \epsilon \rme^{-i \phi_2}}
=\rme^{i(\phi_1-\phi_2)} \frac{1 - \frac12 \epsilon \rme^{-2 i \phi_1}}
{1 - \frac12 \epsilon \rme^{-2 i \phi_2}}
= \rme^{i(\phi_1-\phi_2)}
\frac{(1 - \frac12 \epsilon \rme^{-2 i \phi_1})
(1 - \frac12 \epsilon \rme^{2 i \phi_2})}
{1 + \frac14 \epsilon^2 - \epsilon \cos(\phi_2)}
\end{equation}
The $\rme^{i(\phi_1-\phi_2)}$ is the expected phase measurement,
but the $(1 - \frac12 \epsilon \rme^{-2 i \phi_1})
(1 - \frac12 \epsilon \rme^{2 i \phi_2})$ 
term produces an unsupressed error. This error
is zero when $\phi_2=\phi_1$ or $\phi_2=\phi_1+\pi$, but not in
general.

\begin{figure}
\begin{center}
\psfig{figure=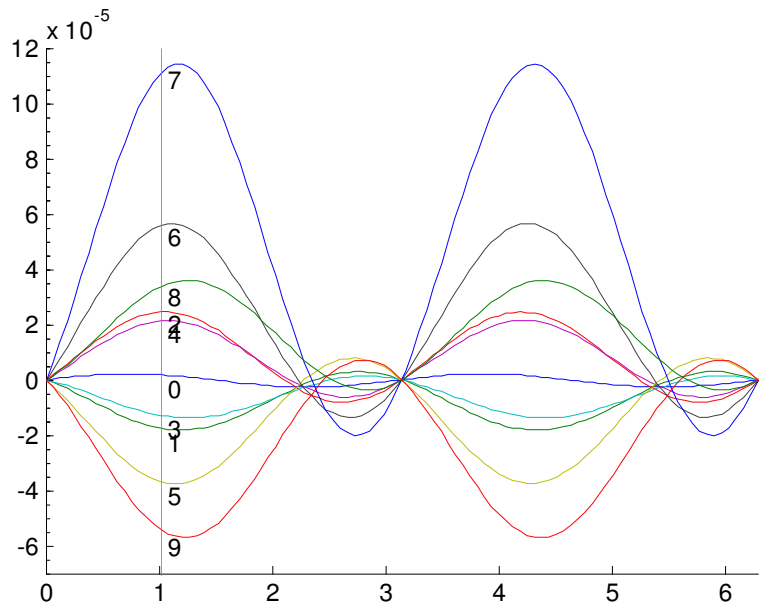,width=10cm}
\end{center}
\caption{\label{fig:cmn}The difference between the measured and
  expected phase difference, as a function of the expected phase
  difference. The different curves show subsequent phase measurements,
  each curve has identical phase noise for different expected phase
  differences.}
\end{figure}

To further illustrate this noise source a numerical simulation of the
gedankenexperiment shown in \Fref{fig:gedanken} has been carried
out. The laser frequency noise has been modelled as:
\begin{equation}
N_f=\cases{1 {{\rm Hz}\over\sqrt{\rm Hz}} & for $f<10^4$ Hz\\
\hbox{falls as 1/f} & for $f>10^4$ Hz}
\end{equation}
as a stabilised NPRO laser, with the knee at $f=10^4$ Hz given by the
bandwidth of the piezo electric control crystal. For such a laser the
phase noise only falls below $10^{-6}\hbox{ cycles}/\sqrt{\rm Hz}$ for
$f\gtrsim 10^5$ Hz. This is suggestive that the
heterodyne frequency 
needs to be in excess of $f=10^5$ Hz. In the simulation the phase has
been measured by 
digitisation of the heterodyne, followed by DFT/FFT, with the various
parameters set as:
\begin{equation}
\label{lparm}
\eqalign{
f_{\rm heterodyne}&=10^5\hbox{Hz}\\
f_{\rm sample}&=1.024\hbox{MHz}\\
{\rm FFT\ length} &= 1024 \\
f_{\rm phase\ measurement}&= 1\hbox{kHz}}
\end{equation}
The simulation results are shown in \Fref{fig:cmn}. To interpret this
figure consider if the properties of optical bench are such that a
phase difference of $\Delta\phi=1$ rad is expected. The measured phase
difference differs for $\Delta\phi=1$ in subsequent phase
measurements as shown on the vertical line and labelled $0,\ldots,9$. The
variation in phase is at level of $\sim 10^{-4}$ rad. As a power
spectral density this corresponds 
$\sim 5\times10^{-7}\hbox{ cycles}/\sqrt{\rm Hz}$, as expected from the
analytic calculation. If the optical bench had a different phase
offset between the two measurements then the magnitude of the phase
error varies, as shown along the $x$ axis of \Fref{fig:cmn} for the
{\it identical} phase noise from the lasers. It can be seen
that there is no phase error when the phase difference is a multiple
of $\pi$.

At first inspection this does not seem to be a difficulty for
LISA, as typical heterodyne frequencies will be in excess of 1MHz,
well beyond the frequency found above. However the problems occur for
the phase measurement parameters \eref{lparm}, where the digitisation
rate of $\sim 1$MHz is beyond that possible with space qualified
hardware. The phase measurement systems considered under the LISA PMS
contract have circumvented this by electrically downconverting the
heterodyne to a lower frequency, before digitisation. However as the
heterodyne is measured at lower frequency than the signal on the
photodiode, the noise that enters at the phase measurement is
sensitive to phase noise at twice the {\it measurement}
frequency. This phase measurement frequency is typically well below
the $\sim 10^5$ Hz frequency at which the laser phase noise becomes
significant.

Similar sources of noise that originate at twice the heterodyne
frequency have independently also been recently identified
\cite{folkner}.

\section{Phasor processing}

The difficulty is caused because a sinusoidal heterodyne lacks the
definition required. Specifically the maximum and minimum of a
sinusoid show little sensitivity to phase, whilst the zero crossing
shows maximal sensitivity to phase. This in turn means that phase
measured from a sinusoid is most sensitive to phase noise during the
zero crossings of the sinusoid. 
The difficulty can be cured if the quadrature component of
the heterodyne can be obtained as well as the in-phase
component. This spreads the phase information evenly in the
signal. Mathematically this means that the full phasor 
representation, including 
imaginary part, can be constructed for the heterodyne:
\begin{equation}
\hskip -2cm
\cos(\omega t + \phi + \epsilon \sin( 2 \omega t))
+ i \sin(\omega t + \phi + \epsilon \sin( 2 \omega t))
= \exp\left(i(\omega t + \phi + \epsilon \sin( 2 \omega t))\right)
\end{equation}
The phasor extracted from this signal has the $\epsilon$ term missing:
\begin{equation}
\exp\left(i(\omega t + \phi + \epsilon \sin( 2 \omega t))\right)
\stackrel{FT}\Rightarrow \rme^{i \phi}
\end{equation}
and hence only the baseband phase noise is present, which is
successfully cancelled via TDI.

So how can the in-phase and quadrature components of the heterodyne be
physically constructed? This is traditionally achieved during
downconversion of the heterodyne from higher frequency, simultaneously
downconverting against both an in-phase and quadrature signals. For
example if the LISA heterodyne is originally at say 20MHz, but is
downconverted to 10kHz for measurement, this downconversion can
introduce the in-phase and quadrature components.

This is not the only downconversion in the LISA laser
interferometry. The interference of the two lasers beams on the
photodiode forms an optical downconversion between the two beams, from
the optical frequency $\sim 3\times 10^{14}{\rm Hz}$ down to the
electrical heterodyne frequency. This down conversion can also be used
to generate in-phase and quadrature signals \cite{digital}, using the
following method: 
\begin{itemize}
\item Using quarter wave plates circularly polarise one beam, and
  linearly polarise the other. The two linear polarisations of the
  circularly polarised light are in quadrature, whilst the two linear
  polarisations (at $45^\circ$) to the linearly polarised light are in
  phase.
\item Separately interfere the two linear polarisations of each
  beam (at $45^\circ$ to the linearly polarised light). This can be
  achieved using the two outputs of a polarising beam splitter.
\item The two heterodynes formed from the two polarisations are in
  quadrature.
\end{itemize}
Comparing the optical to electrical in-phase and quadrature generation
the following comments can be made:
\begin{itemize}
\item The optical method requires a more complicated optical bench
  design.
\item Although the optical method splits the laser beams, this does
  not lead to an increase in shot noise, as both the halves each enter
  in the full phase measurement.
\item The electrical method probably suffers from the problem of phase
  noise at twice the heterodyne. However the relevant frequency is
  twice the heterodyne {\it on the photodiode}. Depending on the level
  of phase noise this may force the heterodyne frequency to an
  unacceptably high value.
\item The optical method also has contributions from phase noise but at
  twice the optical frequency, {\it i.e.} $\sim 6\times 10^{14}{\rm
  Hz}$, at which the laser phase noise will be insignificant.
\end{itemize}
During the study the preference has been for the electrical
generation of the in-phase and quadrature signals. However a full
tradeoff is required before a final decision is made. 

\section{Conclusions}
This ESA funded study by SEA and its team into the Phase Measurement
System for LISA has 
demonstrated many previously unexpected interactions between LISA
systems. These interactions impose additional requirements on the
Phase Measurement System that significantly complicate the design of
LISA. A few of the difficulties have been summarised in this article;
more details can be found in \cite{task3,task4}. None of the difficulties
seem unsurmountable, provided they are 
taken into account sufficiently early in the system detailed design.

\end{document}